\documentclass[superscriptaddress,aps,prl,twocolumn]{revtex4}

\def\beq{\begin{equation}}
\def\eeq{\end{equation}}
\def\ba{\begin{eqnarray}}
\def\ea{\end{eqnarray}}

\usepackage{graphicx}

\begin{document}

\title{Superfluid Fermi Gases with Large Scattering Length}
\stepcounter{mpfootnote}
\author {J. Carlson} 
\affiliation {Theoretical Division, 
  Los Alamos National Laboratory,
  Los Alamos, New Mexico 87545, U.S.A.} 
\author {S-Y Chang}
\affiliation { Department of Physics,  
  University of Illinois at Urbana-Champaign,
        1110 W. Green St., Urbana, IL 61801, U.S.A.}
\author {V. R. Pandharipande}
\affiliation { Department of Physics,  
  University of Illinois at Urbana-Champaign,
        1110 W. Green St., Urbana, IL 61801, U.S.A.}
\author {K. E. Schmidt}
\altaffiliation[Permanent Address:]{
Department of Physics and Astronomy, Arizona State University, Tempe,
AZ 85287, U.S.A.}
\affiliation { Department of Physics,  
  University of Illinois at Urbana-Champaign,
        1110 W. Green St., Urbana, IL 61801, U.S.A.}

\date{\today}

\begin{abstract}

We report quantum Monte Carlo calculations of superfluid Fermi gases 
with short-range two-body attractive interactions with infinite scattering 
length.  The energy of such gases is estimated to be $(0.44 \pm 0.01)$ times 
that of the noninteracting gas, and their pairing gap is approximately
twice the 
energy per particle. 

\end{abstract}

\maketitle

In dilute Fermi gases the pair interactions have a range much smaller 
than the interparticle spacing.  However, when the two-particle scattering length 
is large, these short range interactions can modify the gas properties significantly.  
A well known example is low density neutron matter which may occur in the 
inner crust of neutron stars \cite{pethick1995}.  The two-neutron interaction has 
a range of $\sim 2$ fm, but the scattering length is large, $- 18$ fm, so that even at 
densities as small as one percent of the nuclear density the parameter $ak_F$ 
has magnitude much larger than one.  Bertsch proposed in 1998
that solution of the idealized 
problem of a dilute Fermi gas in the limit $ak_F \rightarrow -\infty$ could 
give useful insights into the properties of low density neutron gas.

Cold dilute gases of $^6$Li atoms have been produced in atom traps. 
The interaction between these atoms can be tuned using a known Feshbach resonance; 
and the estimated value of $ak_F$ in the recent experiment \cite{ohara2002}
is $\sim - 7.4$. 
As the interaction strength is increased beyond that for $a=-\infty$, we get bosonic 
two-fermion bound states.  In this sense a dilute Fermi gas with large $a$ is
in between weak coupling 
BCS superfluid and dilute Bose gases with Bose-Einstein condensation \cite{randeria95}. 
Attempts to produce Bose gases in the limit, $a/r_0 \rightarrow \infty$ 
using Feshbach resonances \cite{stenger1999,roberts2001}, 
are in progress, and their energy has been recently estimated using variational methods 
\cite{cowell02}.   

In the $a \rightarrow -\infty$ 
limit $k_F^2/m$ is the only energy scale,
and the ground state energy of the interacting dilute Fermi gas is proportional to the 
noninteracting Fermi gas energy:
\beq
E_0(\rho)= \xi~E_{FG} = \xi~\frac{3}{10}~\frac{k_F^2}{m}~. 
\eeq
Baker \cite{baker1999} and Heiselberg \cite{heiselberg2001}
have attempted to obtain the 
value of the constant $\xi$ from expansions of 
the Fermi gas energy in powers of $ak_F$.  
Heiselberg obtained $\xi = 0.326 $, while 
Baker's values are $\xi = 0.326$ and $0.568$. 

Fermi gases with attractive pair interaction become superfluid at low 
temperature.
The BCS expressions in terms of the scattering length were given
by Leggett \cite{leggett1980}, and they were used to study 
the properties of superfluid dilute Fermi gases,  
as a function of $ak_F$, by Engelbrecht, Randeria and S{\'a}
de Melo \cite{engelbrecht1997}.
For $ak_F=-\infty$ they obtain an upperbound, $\xi = 0.59$, using 
the BCS wave function.  
These gases are also estimated to 
have large gaps comparable to the ground state energy per 
particle.  

Here we report studies of Fermi gases with quantum Monte Carlo 
methods using the model potential:
\beq
v(r) = - \frac{2}{m} \frac{\mu^2}{\cosh^2(\mu r)}~. 
\eeq
The zero energy solution of the two-body Schr{\"o}dinger equation
with this potential 
is $\tanh(\mu r)/r$ and corresponds to $a=-\infty$.  The effective range is $2/\mu$, 
and in order to ensure that the 
gas is dilute we use $\mu r_0 > 10 $, where $r_0$ is the unit radius; $\rho r_0^3=3/4\pi $. 
%\beq
%\rho = \frac{k_F^3}{3\pi^2} = \frac{3}{4 \pi r_0^3}~. 
%\eeq
All the results presented here are for $\mu r_0 = 12$; however some of the 
calculations were repeated for $\mu r_0 = 24 $ and the results extrapolated to 
$1/\mu \rightarrow 0$.

We have carried out fixed node Green's function 
Monte Carlo \cite{anderson75} (FN-GFMC) calculations with trial wave functions 
of the form:  
\beq
\Psi_V ({\bf R}) = \prod_{i,j'} f(r_{ij'}) \Phi ({\bf R})~, 
\label{eq:wavefun}
\eeq
where $i,j,...$ and $i',j',...$ label spin up and down particles, and the configuration 
vector ${\bf R}$ gives the positions of all the particles.  Only the antiparallel 
spin pairs are correlated in this $\Psi_V$ with the Jastrow function $f(r_{ij'})$.  The 
parallel spin pairs do not feel the short range interaction due to Pauli exclusion. 

In FN-GFMC the $\Psi_V$ is evolved in imaginary time with the operator $e^{-H\tau}$ 
while keeping its nodes fixed to avoid the fermion sign problem.  In the limit 
$\tau \rightarrow \infty$ it yields the lowest energy state with the nodes of 
$\Psi_V$.  These nodes, and hence the FN-GFMC energies, 
do not depend upon the positive definite Jastrow function. 
Nevertheless it is useful to reduce the variance of the FN-GFMC calculation. 
In the present work we use approximate solutions 
of the two-body Schr{\"o}dinger equation:
\beq
\left[- \frac{1}{m} \nabla^2 + v(r)\right]f(r<d)=\lambda f(r<d) ~,
\eeq
with the boundary conditions $ f(r>d)=1$ and $f'(r=d)=0$ \cite{cowell02}. 
The value of $d$ is obtained by minimizing the energy 
calculated using variational Monte Carlo.  Note that a dilute Fermi gas is stable 
even when $a \rightarrow - \infty$, unlike dilute Bose gases in the $a \rightarrow \infty$ 
limit. 

The calculations are carried out in a periodic cubic box having $\rho L^3 = N$. 
The single particle states in this box are plane waves with momenta ${\bf k}_i$:
\beq
{\bf k}_i = \frac{2 \pi}{L}(n_{ix}{\hat x}+n_{iy}{\hat y}+n_{iz}{\hat z})~. 
\eeq
The free-particle energies depend only on $I=n_x^2+n_y^2+n_z^2$.
For $N=14$ and 38 we have closed shells having states with $I\leq 1$ and $I \leq 2$ 
occupied.  The commonly used Jastrow-Slater (JS) $\Psi_V({\bf R})$ is obtained 
by using: 
\beq
\Phi_{\rm S}= [{{\cal A}}\prod_{I<I_{\rm max}}e^{i{\bf k_i}\cdot{\bf r}_j}]
[{{\cal A}}\prod_{I<I_{\rm max}}e^{i{\bf k_i}\cdot{\bf r}_j'}]~, 
\eeq
in Eq. \ref{eq:wavefun}.  The more general, Jastrow-BCS $\Psi_V({\bf R})$ has: 
\beq
\Phi_{\rm BCS} = {{\cal A}}[\phi(r_{11'}) \phi(r_{22'}) ... \phi(r_{nn'})] ~, 
\eeq 
with $n=N/2$.  The antisymmetrizer ${{\cal A}}$ in the $\Phi_{\rm BCS}$ 
separately antisymmetrizes between the spin up and down particles.  
The $\Phi_{\rm BCS}$ describes the component of the BCS state with $N$ particles when
\ba
|{\rm BCS}\rangle &=& \prod_i (u_i + v_i a^{\dagger}_{{\bf k}_i \uparrow}
a^{\dagger}_{{-\bf k}_i \downarrow})|0\rangle ~,  \\ 
\phi(r)&=&\sum_i \frac{v_i}{u_i} e^{i{\bf k}_i\cdot{\bf r}}~. 
\label{eq:pairphi}
\ea
The nodal surfaces of $\Phi_{BCS}$ depend upon the pairing function $\phi(r)$, 
and equal those of $\Phi_S$ when $v_i=0$ for all $k_i > k_F$. 

FN-GFMC gives upperbounds to the energy, which equal 
the exact value when the trial wave function has the nodal structure of the ground state. 
Therefore, we can determine the $\phi(r)$ variationally by minimizing the 
FN-GFMC energy.  We use the parameterization: 
\ba
\phi({\bf r}) &=&\tilde{\beta} (r) + \sum_{i,~I \leq I_C} \alpha_I \exp [ i \bf{k}_i \cdot {\bf r}]~, \\
\tilde{\beta}(r)&=& \beta(r)+\beta(L-r)-2 \beta(L/2)~, \\
\beta (r) &=& [ 1 + \gamma b  r ]\ [ 1 - \exp [ - c b r ]] \frac{\exp [ - b r ]}{c b r}~. 
\ea
The function $\tilde{\beta}(r)$ 
has a range of $L/2$,  the value of $\gamma$ is chosen such that it has 
zero slope at the origin, and $I_C=4$ here.

\begin{figure}
\includegraphics[width=\columnwidth]{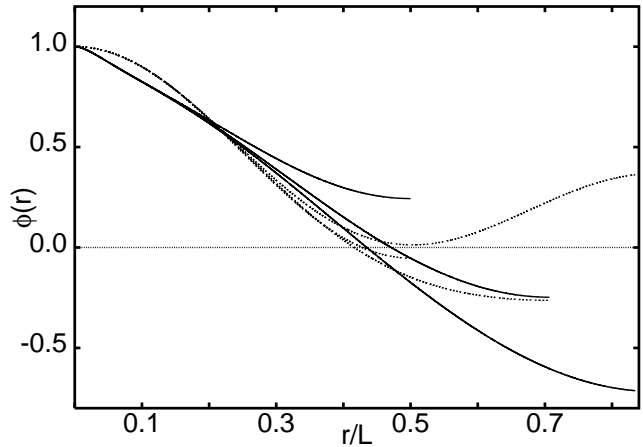}
\caption{ The optimum $\phi({\bf r})$ (solid 
lines) and the $\phi_S$ (dashed lines) for $N=38$.  Curves ending at $L/2$, $L/\sqrt{2}$ 
and $L\sqrt{3/4}$ are in the 001, 011 and 111 directions of the periodic box.
}
\label{phifig}
\end{figure}

\begin{figure}
\includegraphics[height=\columnwidth,angle=270]{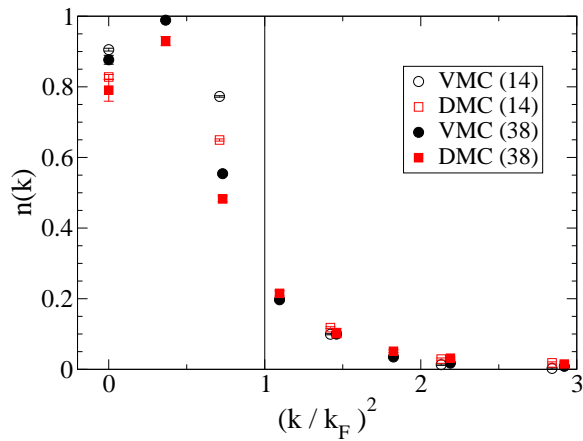}
\caption{The momentum distribution of particles
}
\label{f.kdist}
\end{figure}

The parameters $b, c$ and $\alpha_{I \leq I_C}$ of $\phi({\bf r})$ are optimized by
choosing a random distribution of initial values and measuring the
parameters of the lowest-energy (longest-lasting) configurations in 
FN-GFMC calculation.  
For 38 particles it produces an optimum set of parameters
$br_0=0.8$, $c = 10$, $\alpha_{I=0,4}=0.016, 0.466, 0.0068, 0.00091, 0.0007$
which   
give the smallest FN-GFMC energy having $\xi=0.440(2)$.  Calculations 
in which $\beta(r)=0$ give optimum values 
$\alpha_{I=0,4}=0.24,1.0,0.2,0.057,0.035$ and $\xi=0.459(2)$, while the Slater 
$\phi_S({\bf r})$ having $\beta(r)=0$ and $\alpha_{I=0,4}=1,1,1,0,0$ 
gives a much larger $\xi=0.54$.   

The optimum $\phi({\bf r})$ is compared with the $\phi_S({\bf r})$
in Fig. \ref{phifig}; it has a sharper peak at $r=0$.  This peak depends upon 
the Jastrow function $f(r)$ acting between all the $N^2/4$ antiparallel spin 
pairs.  For example, the $\phi(r)$ obtained by solving the 
BCS equation with the bare potential in uniform gas without the $f(r)$ 
has a much sharper peak. 

The optimum $\phi({\bf r})$ has
$\alpha_0 < \alpha_1$ for N = 38; in contrast the  
variationally determined BCS wave function has $\alpha_0 \geq \alpha_1$.  
The momentum distribution of particles in the trial and evolved 
($\tau=0$ and $\infty$) wave functions are shown in Figure 2.
For N = 38 the occupation of the $I=0$ state is smaller than the $I=1$, 
calculations with much larger values of $N$ are planned to test if this is a 
finite box size effect. 

We have attempted further optimizations by incorporating backflow
\cite{feynman1956,schmidt1981} 
into the BCS pair functions $\tilde{\phi}$. 
Initial calculations indicate that this will reduce the $\xi$ by
$\approx$ 0.02.  On the other hand, estimates of the corrections
due to the finite range of the present interaction indicate
that going to the $1/\mu \rightarrow 0$ limit will raise $\xi$ by a similar amount.
Thus our present upperbound for the constant $\xi$ is 0.44(1). 

In order to estimate the gap $\Delta$ of this superfluid we studied differences between 
energies of systems with odd and even number of particles.  
A general wave function with $n$ pairs,
$u$ spin up and $d$ spin down unpaired particles can be written as: 
\begin{widetext}
\beq
\label{eq.gbcs}
\Phi_{\rm BCS}(R) =  {{\cal A}} \left \{ 
\left[ \phi(r_{11'})...\phi(r_{nn'})\right] \left [
\psi_{1\uparrow}({\bf r}_{n+1})...\psi_{u\uparrow}({\bf r}_{n+u})\right ]
\left[\psi_{1\downarrow}({\bf r}_{(n+1)'})...\psi_{d\downarrow}({\bf r}_{(n+d)'})\right ]
\right \} ~. 
\eeq
The unpaired particles are in $\psi_{i\uparrow}$ and $\psi_{j\downarrow}$ single 
particle states.  
We can write this wave function as the determinant of an
$M\times M$ matrix where $M= n+u+d $ \cite{bouchaud1988}.
For example, when $u=2$ and $d=3$ the 
matrix is given by: 
\begin{equation}
\left (
\begin{array}{cccccc}
\phi(r_{11'}) & \phi(r_{12'}) & ... &
\phi(r_{1(n+d)'}) &
\psi_{1\uparrow}({\bf r}_1) & \psi_{2\uparrow}({\bf r}_1) \\
\phi(r_{21'}) & \phi(r_{22'}) & ... &
\phi(r_{2(n+d)'}) &
\psi_{1\uparrow}({\bf r}_2) & \psi_{2\uparrow}({\bf r}_2) \\
\vdots & \vdots & \vdots\vdots\vdots & \vdots & \vdots & \vdots\\
\phi(r_{(n+u)1'}) & \phi(r_{(n+u)2'}) & ... &
\phi(r_{(n+u)(n+d)'}) &
\psi_{1\uparrow}({\bf r}_{n+u}) & \psi_{2\uparrow}({\bf r}_{n+u}) \\
\psi_{1\downarrow}({\bf r}_{1'}) &
\psi_{1\downarrow}({\bf r}_{2'}) & ... &
\psi_{1\downarrow}({\bf r}_{(n+d)'}) &  0 & 0 \\
\psi_{2\downarrow}({\bf r}_{1'}) &
\psi_{2\downarrow}({\bf r}_{2'}) & ... &
\psi_{2\downarrow}({\bf r}_{(n+d)'}) &  0 & 0 \\
\psi_{3\downarrow}({\bf r}_{1'}) &
\psi_{3\downarrow}({\bf r}_{2'}) & ... &
\psi_{3\downarrow}({\bf r}_{(n+d)'}) &  0 & 0 \\
\end{array}
\right )
\end{equation}
\end{widetext}
The fact that the general $\Phi_{\rm BCS}({\bf R})$ can be expressed as a determinant 
makes it possible to perform numerical calculations for large values of $N$.  When 
$N=2n$, the fully paired ground state has $u=d=0$, while those  
of systems with $N=2n+1$ have either $u$ or $d=1$.  

The FN-GFMC ground state energies for various values of $N$ are shown in
figure 3.
The straight dotted line in Fig. 3 is $0.44 E_{FG}$.  
The calculated energies have the odd-even gap expected in superfluids, and 
well known in nuclei.  The values of the odd-even gap:
\beq
\Delta(N=2n+1)=E(N)-\frac{1}{2}(E(N-1)+E(N+1))~, 
\eeq
are shown in Fig. 4. 
The estimated value of the gap is $\sim 0.9 E_{FG}$ or $\sim 2\xi E_{FG}$.
In fact the odd particle removal energies, $E(N=2n+1)-E(N=2n)$, at fixed density, 
are $\sim (4/3)E_{FG}$.  The 
odd particles in the interacting gas have energies higher than that for noninteracting. 
Apparently the odd particles do not gain any benefit from the attractive pair potential, on the 
other hand they hinder the pairing of the others.  BCS calculations including polarization 
correction \cite{gorkov61,heiselberg2000} give $\Delta=0.81 E_{FG}$ in the large $a$ limit.

\begin{figure}
\includegraphics[width=\columnwidth]{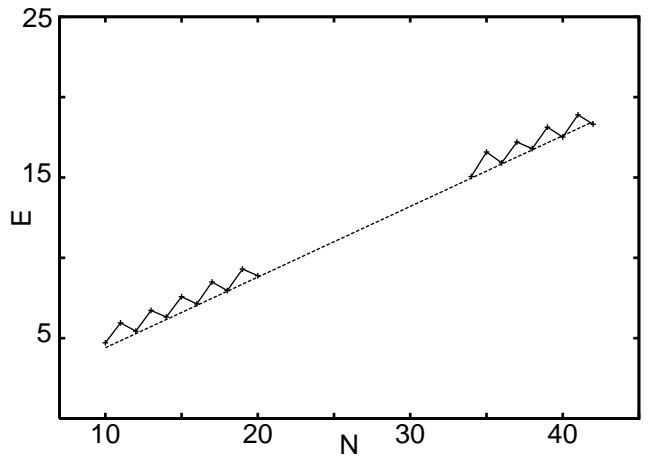}
\caption{ The $E(N)$ in units of $E_{FG}$
}
\label{f.energy}
\end{figure}

\begin{figure}
\includegraphics[width=\columnwidth]{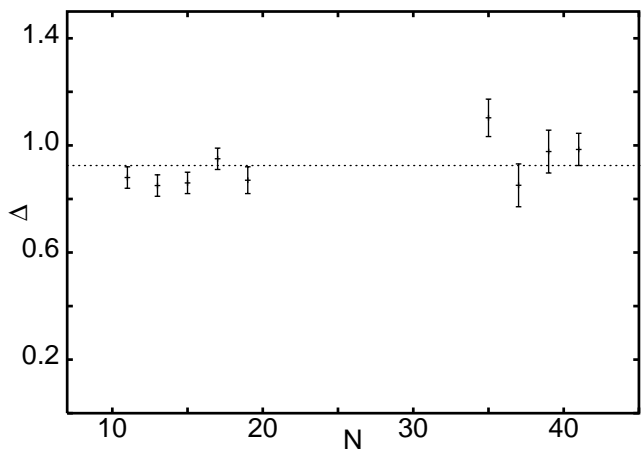}
\caption{
The gap in units of $E_{FG}$
}
\label{f.gap}
\end{figure}

Several consequences of the strong pairing in this superfluid gas are seen in 
the calculated energies.  Noninteracting Fermi gases have shell gaps at $N=14$ 
and 38; they are not noticeable in this gas.  
The ground states of 15 and 17 particle systems have momenta with $I=1$ 
rather than the $I=2$ in noninteracting states and the $I=0$
expected in the limit of strongly-bound pairs.

Some of the differences between the nodal structures of the JS and
J-BCS wave functions can be seen by considering the case where
${\bf r}_i={\bf r}_{i'}$. For the JS case, the up and down determinants
will then be identical and the complete wave function will be the
square of one of these determinants. We now imagine exchanging the
positions of two pairs by rotating them around their center of mass.
Since each determinant must change sign, the JS wave function 
must go through zero during this exchange. 
When the pairs are separated by small distances
the up and down determinants are no longer equal. Thus they will change signs 
at different points along the exchange path. We therefore expect a negative 
region which will effectively
block these ``two-boson'' exchanges for fixed node calculations.
In the J-BCS case, the exchanges can occur without crossing a node.
In the composite boson limit where $\phi(r)$ is strongly peaked around
the origin, there is no sign change under pair exchanges  when 
all the pairs are well separated.

In order to further
understand the difference between the JS and J-BCS wave functions we studied 
their nodal structure for the following three-pair exchange.  
In randomly chosen 
configurations distributed with $\Psi_{\rm J-BCS}^2({\bf R})$ the three closest pairs $ii'$, 
$jj'$ and $kk'$ were identified.  Their center of masses are denoted by ${\bf S}_l$. 
The wave functions $\Psi'(x)$ are calculated for the positions defined as follows:
All particles $m,m' \neq i,i',j,j',k,k'$ retain their positions in the random 
configuration.  The positions of $i,j,k$ are given by: 
\beq
{\bf r}_i = {\bf S}_i + {\bf s} + x({\bf S}_j - {\bf S}_i ) ~, 
\eeq
and cyclic permutations of it.  Here ${\bf s}$ is the relative distance between particles 
in a pair.  Those of $i',j',k'$ have $-{\bf s}$ in place of ${\bf s}$, 
and the typical value, $|{\bf s}|=0.25 r_0$ is used in these studies.  
The three pairs complete a circular exchange $ii' \rightarrow jj' 
\rightarrow kk'$, in the $x=0$ to 1 
interval.  We calculate the ratio $\Psi'(x)/\Psi'(x=0)$ for many configurations.  Note that 
$\Psi'(1)/\Psi'(0) = 1$.  In a fixed node calculation the space where this ratio is negative 
is blocked for the diffusion of the configuration.  For JS and J-BCS wave functions the ratios 
are negative, on average, over 29 and 17 \% of the $x=0$ to 1 domain.  For about half
of the configurations 
the J-BCS had positive ratio for all values of $x$, while only 20 \% of
the JS configurations have this property.

We therefore picture the change in the nodal structure in going from
the JS to the J-BCS wave functions as an opening up of the configuration space to
allow pairs to exchange without crossing a node. 
%Assuming that the number
%of nodal regions does not change, this increase in volume for some degrees
%of freedom must be matched by decreases in volume for others. 
For systems with a paired ground state, the J-BCS presumably allows off diagonal long range
order via these pair exchanges.
In most cases the energy difference between the normal state evolved from the JS 
wave function and the superfluid state evolved from J-BCS is very small $(< 0.1 \%)$, 
and calculations of the type presented here are difficult.  However, in dilute
Fermi gases  
with large negative $a$ this difference is $\sim$ 20 \% and calculations of the 
superfluid are possible with bare forces. 

We would like to thank M. Randeria, A. J. Leggett, and D.G. Ravenhall for
useful discussions.  The work of JC is supported 
by the US Department of Energy under contract W-7405-ENG-36, while that of 
SYC, VRP and KES is partly supported 
by the US National Science Foundation via grant PHY 00-98353.

\end{document}